\documentclass[aps,twocolumn,showpacs]{revtex4-2}
\usepackage{amsmath}
\usepackage{epsfig}
\usepackage{booktabs} %调整表格线与上下内容的间隔
\usepackage{multirow}
\usepackage{graphicx} 
\usepackage{subcaption}
\usepackage{caption}
\begin{document}

\title{``Inevitable'' nonstrange tetraquark states and $\rho\rho$ interaction}

\author{Yanyue Pan$^1$}\email[E-mail: ]{251002046@njnu.edu.cn}
\author{Yuying Hao$^1$}\email[E-mail: ]{231002023@njnu.edu.cn}
\author{Shirui Zhao$^1$}\email[E-mail: ]{251002076@njnu.edu.cn}
\author{Hongxia Huang$^1$}\email[E-mail: ]{hxhuang@njnu.edu.cn (Corresponding author)}
\author{Jialun Ping$^1$}\email[E-mail: ]{jlping@njnu.edu.cn (Corresponding author)}
\affiliation{$^1$Department of Physics, Nanjing Normal University, Nanjing, Jiangsu 210097, China}

\begin{abstract}
Investigation of the interaction between two $\rho$ mesons with $IJ = 00,~02,~11,~20$ and $22$ is conducted systematically in the framework of the quark delocalization color screening model. The results show that there is a strong attractive potential between two $\rho$ mesons, which makes it easier for these two mesons to form bound states. The binding energies vary from a few MeV to over a hundred MeV. Moreover, the results of intercluster distances favor a molecular interpretation of these states. The interaction analysis reveals that the $\rho\rho$ and $\Delta\Delta$ interactions share certain similarities. 
%Inspired by the prediction of the dibaryon $d^*$, and based on the analysis of the decay properties, we suppose that $\rho\rho$ with $IJ=02$ may be the most promising candidate of the ``inevitable'' nonstrange tetraquark state, which is worth searching for experimentally. 
Drawing inspiration from the $d^*$ dibaryon prediction and from decay-property analyses, we predict that the $\rho\rho$ with $IJ=02$ and $22$ are the most promising candidates of the ``inevitable'' nonstrange tetraquark states, which deserves experimental search.

\end{abstract}

%\pacs{13.75.Cs, 12.39.Pn, 12.39.Jh}

\maketitle

\setcounter{totalnumber}{5}

\section{\label{sec:introduction}Introduction}

The study of tetraquark dates back to Jaffe in 1977, where he pointed out that certain known $0^+$ mesons $(\epsilon(700), S^*, \delta, \kappa)$ are not conventional $Q\bar{Q}$ states but rather the lightest cryptoexotic $Q^2\bar{Q}^2$ nonet~\cite{Jaffe1,Jaffe2}. 
The renewed interest in tetraquark systems was triggered in 2003 by the successive discoveries of several charmed tetraquarks, such as the $X(3872)$ observed by the Belle Collaboration in $B^{\pm} \rightarrow K^{\pm} \pi^{+} \pi^{-} J / \psi$~\cite{X3872}, the $D_s(2317)$ reported by the BaBar Collaboration in the $D_s^+\pi^0$ invariant mass spectrum~\cite{Ds2317}, the charged charmonium-like state $Z_c(3900)^{\pm}$ reported by BESIII in $e^{+}e^{-} \rightarrow \pi^{+} \pi^{-} J / \psi$~\cite{Zc3900}, and the open-charm exotic structures $X_0(2900)$ and $X_1(2900)$ observed by LHCb in $B^{+} \rightarrow D^{+} D^{-} K^{+}$~\cite{X2900}.

While much of the interest has focused on charmed tetraquarks, fully light tetraquarks exist as well and have likewise been the subject of extensive study. In Ref.~\cite{Lodha}, the mass spectra and decay properties of all light tetraquarks were investigated with both semi-relativistic and non-relativistic frameworks.
Using a dispersive analysis of Roy equations, Caprini \textit{et al.} obtained the $\sigma$ pole at $m_{\sigma}=441-i272$ MeV, establishing its resonant nature, and determined the $f_{0}(980)$ pole within the same framework~\cite{Caprini}. Subsequently a dispersive analysis of $\pi\pi$ scattering data using the GKPY equations determined the pole positions of both the $f_{0}(600)$ and $f_{0}(980)$ at $457-i279$ MeV and $996-i25$ MeV, respectively~\cite{Ruiz,Ruiz2}, which have since served as benchmark inputs for later theoretical studies. Agaev \textit{et al.} discussed the structure, mixing angle, and coupling constants of $f_{0}(500)$ and $f_{0}(980)$ within QCD sum rules~\cite{Agaev}, with results consistent with a tetraquark-dominated picture, while Weinstein and Isgur interpreted $f_{0}(980)$ and $a_{0}(980)$ as weakly bound $K\bar{K}$ molecular states in a nonrelativistic potential model~\cite{Weinstein}.
In 1995, the Crystal Barrel Collaboration confirmed $f_{0}(1370)$ in $\bar{p}p$ annihilation at rest from  LEAR~\cite{f1370}. A prominent signal of $f_{0}(1370)$ was later seen by the BESII Collaboration in $J / \psi \rightarrow \phi \pi^{+} \pi^{-}$ and the resonance was confirmed by the STAR Collaboration in the $\pi^{+} \pi^{-}$ invariant mass spectrum in $pp$ collisions at $\sqrt{s}=200$ GeV~\cite{f1370BES,f1370STAR}. A model-independent dispersive analysis of $\pi\pi$ scattering data has furthermore established the corresponding pole at $1245-i300$ MeV~\cite{Pelaez}. Its branching ratios, however, remain undetermined, its experimental Breit-Wigner width lies between 200 and 500 MeV~\cite{f1370PDG}, and its internal structure is still under considerable debate. Large-$N_c$ QCD study suggests a possible interpretation in which $f_{0}(1370)$, $a_{0}(1450)$, $k_{0}^*(1430)$ and $f_{0}(1500)$ form the lowest tetraquark nonet~\cite{f1370Nc}. 
%In a recent study of Ref.~\cite{f1370kkbar}, $f_{0}(1370)$ was assumed to be a $\kappa\bar{\kappa}$ molecular state, and the partial widths of its main decay channels, including $K\bar{K}$, $\pi\pi$, $\eta\eta$, $4\pi$, and $K\bar{K}\pi\pi$, were calculated. This yielded the decay pattern $\Gamma(K\bar{K}) > \Gamma(4\pi) \approx \Gamma(\pi\pi)$. It was also pointed out that although the existing experimental data differ from the model predictions in certain ratios, the $\kappa\bar{\kappa}$ molecular configuration of $f_{0}(1370)$ cannot yet be ruled out, since most of its experimental data suffer from strong model dependence and mutual inconsistencies.

%In 1995, the Crystal Barrel Collaboration confirmed $f_{0}(1370)$ in $p\bar{p}$ annihilation at LEAR\cite{f1370}, sparking extensive studies on light scalar mesons. 

Our group has also carried out a systematic study of the interactions between light mesons within the quark delocalization color screening model(QDCSM). We found that, for most physical channels, the interaction between two pseudoscalar mesons is repulsive; the one between a pseudoscalar meson and a vector meson is repulsive or weakly attractive; and the one between two vector mesons is attractive~\cite{WuYH1}. The greater likelihood of the attraction between two vector mesons indicates a higher possibility of forming bound states, which is of particular importance for tetraquark studies. 

Additionally, a similar behavior of the interaction properties has been found in hexaquark systems. In Ref.~\cite{baryon}, dibaryon candidates in the $u$, $d$ and $s$ flavor sectors were systematically investigated within the QDCSM, and the results showed that octet-octet interactions are mostly repulsive or weakly attractive, decuplet-octet interactions are always weakly attractive and decuplet-decuplet interactions are always strongly attractive. One of the most representative examples of the decuplet-decuplet interactions is the $\Delta\Delta$ with $IJ=03$ predicted by T.~Goldman \textit{et al.} in 1989. They called the state a nonstrange ``inevitable'' state and named it $d^*$~\cite{d*}. 
Other groups have also investigated the $\Delta\Delta$ interaction and found it to be attractive. The binding energies have been calculated in several works, with values ranging from a few MeV to several hundred MeV~\cite{Oka80,Maltman84,Yuan99,Pang02,Valcarce05,Huang,Gal16}.
%Oka and Yazaki found an attractive force in the ${}^7S_3((S,T)=(3,0))$ state of $\Delta\Delta$, which is large enough to form a bound state with the binding energy of $6.6$ MeV\cite{Oka80}. Using a QCD-like potential model, Maltman found that hidden-color mixing can lead to a much deeper bound state in the $I=0$, $S=3$ channel, with a binding energy of about $260$ MeV relative to the $\Delta\Delta$ threshold\cite{Maltman84}. \cite{Huang}.
Experimentally, a pronounced resonance structure was observed in the $pn \rightarrow d\pi^0\pi^0$ cross section with the resonance mass $M=2.37$ GeV and the width $\Gamma \approx 70$ MeV by the CELSIUS-WASA Collaboration~\cite{WASA08} and WASA-at-COSY Collaboration~\cite{WASA11,WASA14}. %Subsequently, a resonance pole at $(2380\pm 10 - i40 \pm 5)$ MeV was identified in the ${}^3D_3 - {}^3G_3$ coupled partial waves, establishing $d^*$ as a genuine $s$-channel resonance for the first time in 2014\cite{WASA14}.

Motivated by the similarity between the interaction properties of tetraquarks and hexaquarks, we wonder whether a $d^*$-like resonance, well established in decuplet-decuplet systems, also exists in vector-vector meson systems. Since our previous work revealed a deep attraction between two $\rho$ mesons~\cite{WuYH1}, the central aim of present work is to determine whether $\rho\rho$ can form such a nonstrange ``inevitable'' tetraquark state.
 
The structure of this paper is as follows. The information of the quark model and the calculation methods are given in Section II. The numerical results and corresponding discussion are presented in Section III. The last section is a summary.

\section{QUARK MODEL AND CALCULATION METHODS}

\subsection{The quark delocalization color screening model (QDCSM)}
In this work, we employ the quark delocalization color screening model (QDCSM), which has been explored in detail in the  Refs.~\cite{QDCSM0,QDCSM1}. The explicit form of the Hamiltonian is given as follows. 

\begin{widetext}
\begin{eqnarray}
H & = & \sum_{i=1}^4 \left( m_i+\frac{p_i^2}{2m_i}\right)-T_{CM} +\sum_{j>i=1}^4
\left(V_{ij}^{CON}+V_{ij}^{OGE}+V_{ij}^{OBE} \right),\\[3pt]
V_{ij}^{CON} & = & \left \{
\begin{array}{ll}
-a_{c}\boldsymbol {\mathbf{\lambda}}^c_{i}\cdot
\boldsymbol{\mathbf{ \lambda}}^c_{j}~\left(r_{ij}^2+\nu_{0}\right),&
   \mbox{if \textit{i},\textit{j} in the same cluster}\\[6pt]
-a_{c}\boldsymbol {\mathbf{\lambda}}^c_{i}\cdot
\boldsymbol{\mathbf{\lambda}}^c_{j}~(
\frac{1-e^{-\mu_{ij}\mathbf{r}_{ij}^2}}{\mu_{ij}}+\nu_{0}),& \mbox{otherwise}
\end{array}
\right.\label{QDCSM-vc}\\[3pt]
V^{OGE}_{ij} & = & \frac{1}{4}\alpha_s \boldsymbol{\lambda}^{c}_i \cdot
\boldsymbol{\lambda}^{c}_j
\left[\frac{1}{r_{ij}}-\frac{\pi}{2}\delta(\boldsymbol{r}_{ij})(\frac{1}{m^2_i}+\frac{1}{m^2_j}
+\frac{4\boldsymbol{\sigma}_i\cdot\boldsymbol{\sigma}_j}{3m_im_j})-\frac{3}{4m_im_jr^3_{ij}}
S_{ij}\right] \label{sala-vG} \\[3pt]
V^{OBE}_{ij} & = & V_{\pi}( r_{ij})\sum_{a=1}^3\lambda
_{i}^{a}\cdot \lambda
_{j}^{a}+V_{K}(r_{ij})\sum_{a=4}^7\lambda
_{i}^{a}\cdot \lambda _{j}^{a}
+V_{\eta}(r_{ij})\left[\left(\lambda _{i}^{8}\cdot
\lambda _{j}^{8}\right)\cos\theta_P-(\lambda _{i}^{0}\cdot
\lambda_{j}^{0}) \sin\theta_P\right] \label{sala-Vchi1} \\[3pt]
V_{\chi}(r_{ij}) & = & {\frac{g_{ch}^{2}}{{4\pi
}}}{\frac{m_{\chi}^{2}}{{\
12m_{i}m_{j}}}}{\frac{\Lambda _{\chi}^{2}}{{\Lambda _{\chi}^{2}-m_{\chi}^{2}}}}%
m_{\chi} \left\{(\boldsymbol{\sigma}_{i}\cdot
\boldsymbol{\sigma}_{j})
\left[ Y(m_{\chi}\,r_{ij})-{\frac{\Lambda_{\chi}^{3}}{m_{\chi}^{3}}}%
Y(\Lambda _{\chi}\,r_{ij})\right] \right.\nonumber \\
&& \left. +\left[H(m_{\chi}
r_{ij})-\frac{\Lambda_{\chi}^3}{m_{\chi}^3}
H(\Lambda_{\chi} r_{ij})\right] S_{ij} \right\}, ~~~~~~\chi=\pi, K, \eta, \\[3pt]
S_{ij}&=&\left\{ 3\frac{(\boldsymbol{\sigma}_i
\cdot\boldsymbol{r}_{ij}) (\boldsymbol{\sigma}_j\cdot
\boldsymbol{r}_{ij})}{r_{ij}^2}-\boldsymbol{\sigma}_i \cdot
\boldsymbol{\sigma}_j\right\},\\[3pt]
H(x)&=&(1+3/x+3/x^{2})Y(x),~~~~~~
 Y(x) =e^{-x}/x. \label{sala-vchi2}
\end{eqnarray}
\end{widetext}

Where $V_{ij}^{CON}$, $V_{ij}^{OGE}$ and $V_{ij}^{OBE}$ denote the confinement potential, the one-gluon exchange potential and the Goldstone boson exchange potential, respectively. $T_{CM}$ represents the kinetic energy of the center of mass; $\alpha_s$ is the quark-gluon coupling constant; $Y(x)$ and $H(x)$ are standard Yukawa functions; $S_{ij}$ is quark tensor operator, which gives no contribution here as only $S$-wave states are involved in this work. 
The coupling constant $g_{ch}$ for the chiral field is determined from the $NN\pi$ coupling constant through
\begin{equation}
\frac{g_{ch}^2}{4\pi}=\left(\frac{3}{5}\right)^2\frac{g_{\pi NN}^2}{4\pi}\frac{m_{u,d}^2}{m_N^2}.
\end{equation}

The color screening parameter $\mu_{ij}$, which is fixed by fitting to the deuteron properties, $NN$ phase shifts and scattering phase shifts, takes the value $\mu_{uu}=0.45~fm^{-2}$.%, $\mu_{us}=0.19~fm^{-2}$, $\mu_{ss}=0.08~fm^{-2}$, satisfying the relation, $\mu_{us}^2=\mu_{uu}\mu_{ss}$.

Other symbols in the above expressions have their usual meanings. In this work, we use the model parameters determined by fitting the meson spectrum, as taken from our previous study on the tetraquark states~\cite{XueY}.

\subsection{The calculation methods}
The resonating group method (RGM) is employed here to study a bound state or a scattering problem. We express the wave function of the tetraquark state as
\begin{equation}
\Psi_{4q} = \mathcal{A} \left\{ [\Psi_A \Psi_B]^{[\sigma]IS} \otimes \chi(R) \otimes \Phi_C(R_C) \right\}^J,
\end{equation}
where the symbol ${\cal A }$ is the anti-symmetrization operator, with the SU(3) symmetry written as $\mathcal{A} = 1 - P_{13} - P_{24} + P_{13}P_{24}$. $[\sigma]=[222]$ gives the total color symmetry, $I$ represents isospin quantum number and $S$ refers to the spin quantum number. $\chi(R)$ is the relative motion wave function of the two clusters, with $R$  being the relative coordinate between them; $\Phi_C(R_C)$ denotes the center-of-mass wave function. $\Psi_{A}$ and $\Psi_{B}$ are the $q\bar{q}$ cluster wave functions,
\begin{equation}
\Psi_{A}=(\frac{1}{2\pi b^{2}})^{3/4}e^{-\boldsymbol{\mathbf{\rho}}^{2}_{A}/(4b^{2})}\eta _{I_{A}S_{A}}\chi_{A}^{c},
\end{equation}

\begin{equation}
\Psi_{B}=(\frac{1}{2\pi b^{2}})^{3/4}e^{-\boldsymbol{\mathbf{\rho}}_{B}^{2}/(4b^{2})}\eta _{I_{B}S_{B}}\chi_{B}^{c},
\end{equation}
%where $\eta _{I_{A}S_{A}}$($\eta _{I_{B}S_{B}}$) represents the multiplied wave functions of flavor and spin of the cluster A(B). $\chi_{A}^{c}$($\chi_{B}^{c}$) is the internal color wave functions of cluster A(B). 
The $\chi(R)$ is obtained by solving the Schr\"odinger equation within the variational approach:
\begin{equation}
\langle \delta \Psi'' | H - E | \Psi' \rangle = 0.
\end{equation}

After performing the variation, the RGM equation is as follows
\begin{equation}
\hspace*{-0.2cm} \int H(\boldsymbol{\mathbf{R''}},\boldsymbol{\mathbf{R'}})\chi(\boldsymbol{\mathbf{R'}})d\boldsymbol{\mathbf{R'}}=E\int N(\boldsymbol{\mathbf{R''}},\boldsymbol{\mathbf{R'}})\chi(\boldsymbol{\mathbf{R'}})d\boldsymbol{\mathbf{R'}},
\end{equation}
where $H(\boldsymbol{\mathbf{R''}},\boldsymbol{\mathbf{R'}})$ and $N(\boldsymbol{\mathbf{R''}},\boldsymbol{\mathbf{R'}})$ are Hamiltonian and norm kernels.

Then the energy $E$ and the wave function can be obtained by solving the RGM equation. In practice, however, direct implementation of the RGM expression is computationally inconvenient. To recast the differential-integral equation of RGM into an algebraic form that can be readily solved numerically, we expand the relative-motion wave function as a series of known Gaussian wave functions:
\begin{eqnarray}
& & \chi(\boldsymbol{R}) = \frac{1}{\sqrt{4\pi}}\sum_{L}(\frac{1}{\pi b^2})^{\frac{3}{4}} \sum_{i} C_{i,L}  \nonumber \\
&& ~~~~\times  \int e^{-\frac{1}{2}(\boldsymbol{R}-\boldsymbol{S}_{i})^{2}/b^{2}} Y_{LM}(\hat{\boldsymbol{S}_{i}})d\Omega_{S_{i}}, ~~~~~
\end{eqnarray}
where $\mathbf{S}_{i}$ is the generating coordinate in the model, denoting the separation of two reference centers. The center of mass motion wave function is
\begin{equation}
\Phi_{C}(\boldsymbol{R}_{C})=(\frac{4}{\pi b^{2}})^{3/4} e^{-2\boldsymbol{R}^{2}_{C}/b^{2}}.
\end{equation}

As for the orbital wave functions, $\phi_{\alpha}(\boldsymbol{S}_{i})$ and $\phi_{\beta}(-\boldsymbol{S}_{i})$ are the single-particle orbital wave functions with different reference centers:
\begin{eqnarray}
\phi_{\alpha}(\boldsymbol{S}_{i})=(\frac{1}{\pi b^{2}})^{3/4}e^{-\frac{1}{2}(\boldsymbol{r}_{\alpha}-\boldsymbol{S}_{i}/2)^{2}/b^{2}},\nonumber\\
\phi_{\beta}(\boldsymbol{-S}_{i})=(\frac{1}{\pi b^{2}})^{3/4}e^{-\frac{1}{2}(\boldsymbol{r}_{\beta}+\boldsymbol{S}_{i}/2)^{2}/b^{2}}.
\end{eqnarray}

Then equation (9) can be written as 
\begin{eqnarray}
\Psi_{4q}= {\cal A} \sum_{i,L} C_{i,L}
\int\frac{d\Omega_{S_{i}}}{\sqrt{4\pi}}\prod^{2}_{\alpha=1}\phi_{\alpha}(\boldsymbol{S}_{i})\prod^{4}_{\beta=3}\phi_{\beta}(-\boldsymbol{S}_{i})\nonumber\\
\times[[\eta_{I_{A}S_{A}}\eta_{I_{B}S_{B}}]^{IS}Y_{LM}(\boldsymbol{S}_{i})]^{J}[\chi_{A}^{c}\chi_{B}^{c}]^{[\sigma]}.~~
\end{eqnarray}

%Since we explor $S$-wave systems $(L = 0)$ only in this work, 
The RGM expression is finally given as follows
\begin{equation}
\sum_{j}C_{j}H_{i,j}=E\sum_{j}C_{j}N_{i,j},
\end{equation}
where $N_{i,j}$ and $H_{i,j}$ are the overlaps and Hamiltonian matrix elements, respectively. In this work, only $S$-wave systems are explored. By solving Eq.~(18), the energies and wave functions of tetraquark states can be obtained.

\section{The result and discussion}
In this work, we first calculate the binding energies of the $S$-wave $\rho\rho$ states with all possible combinations of isospin $I$ and total angular momentum $J$. The results are listed in Table~\ref{binding}.
The binding energy is given by $B_{sc} = E_{sc} - E_{th}$, where $E_{sc}$ represents the single channel energy, and $E_{th}$ is the corresponding theoretical threshold. 
\begin{table}[ht]
\centering
\caption{The binding energies of the $S$-wave $\rho\rho$ states.}
\label{binding}
\setlength{\tabcolsep}{8pt}   % 默认是 6pt
\begin{tabular}{ccccc}
\hline \hline
$IJ$ & $E_{th}$ (MeV) & $E_{sc}$ (MeV) & $B_{sc}$ (MeV) \\ \hline
00 & \multirow{5}{*}{1544.93} & $1420.00$ & $-124.93$ &  \\
02 & & $1457.50$ & $-87.43$ &  \\
11 & & $1475.29$ & $-69.64$ \\
20 & & $1542.62$ & $-2.31$ &  \\
22 & & $1454.51$ & $-90.42$ &  \\
\hline \hline
\end{tabular}
\end{table}

It is evident from Table~\ref{binding} that $\rho\rho$ sustains bound states with all allowed $IJ$ numbers and the binding energies vary significantly with different $IJ$. The largest binding energy is observed with $IJ = 00$, reaching $-124.93$ MeV, followed by $IJ = 22$, $IJ=02$ and $IJ = 11$, with the values of $-90.42$ MeV, $-87.43$ MeV and $-69.64$ MeV, respectively. A significant reduction in binding is observed for $IJ = 20$, which gives only $-2.31$ MeV, the weakest of all $IJ$ states. 
Notably, studies based on chiral effective field theory have employed the $\rho\rho$ state with $IJ=00$ to account for the observed $f_0(1370)$~\cite{Molina08,Geng09,Gulmez17}. Reference~\cite{DML18} employed an improved unitarization method to study the interactions between vector mesons and pointed out that the $\rho\rho$ state with $IJ=00$ is significant to form the scalar meson $f_0(1370)$ or $f_0(1500)$.
In present work, the single channel energy of $IJ=00$ is $1420.00$ MeV, which is closer to $1370$ MeV and may be used to explain the $f_0(1370)$. Besides, some studies associate the $IJ=02$ $\rho\rho$ state with the observed $f_2(1270)$~\cite{Molina08,Geng09}, while others suggest that it is unbound~\cite{Gulmez17,DML18}. To determine whether these $\rho\rho$ states can account for the observed $f_0$ or $f_2$ states, the channel coupling effects should be taken into account and the studies of the properties of the resonance states as well as the decay widths are needed.

%The variation in binding energy among these $IJ$ channels is consistent with the behavior of the $\rho\rho$ interaction explored in Ref.~\cite{WuYH1}, with the effective potential for $IJ=00$ and $IJ=02$ being the deepest and that for $IJ=20$ the shallowest. 
%In addition, Table~\ref{binding} shows that, for a given angular momentum $J$, the binding energy is larger with a smaller isospin $I$. For example, when $J=0$, the binding energy with $I=0$ is larger than that with $I=2$; the same trend persists with $J=2$. Conversely, for a fixed isospin $I$, a larger angular momentum $J$ tends to give a larger binding energy. For instance, the binding energy with $J=2$ is larger than that with $J=0$ for both $I=0$ and $I=2$. In general, a larger binding energy can be observed in the $\rho\rho$ states with smaller isospin and larger total angular momentum.

The distance between the two clusters is calculated to further explore the structure of the $\rho\rho$. For $IJ=00$, $IJ=02$, $IJ=11$, $IJ=20$ and $IJ=22$ the distances are $1.315$ fm, $1.333$ fm, $1.344$ fm, $1.714$ fm and $1.315$ fm, respectively. The assumed radius $b$ of a single cluster is $0.518$ fm. From the above data, it is clear that the distance between the two clusters exceeds $2b$ $(2b=1.036 \mathrm{fm})$, so we assume that $\rho\rho$ tends to form molecular states for all $IJ$ channels under consideration.

To understand why $\rho\rho$ can form bound states, we investigate the contribution of each interaction term to the system with different $IJ$. The results are shown in Fig.~1, where the kinetic energy $(V_{\nu k})$, the confinement $(V_{con})$, the one-gluon exchange $(V_{oge})$, and the one-boson exchange $(V_{\pi},V_{K},V_{\eta})$ terms are presented. It should be noted that $V_{K}$ makes no contribution to $\rho\rho$ studied here, hence, only $V_{\pi}$ and $V_{\eta}$ are included in Fig.~1.
%The effective potential is defined as the following form:
%\begin{equation}
%V(S_{i}) = E(S_{i}) - E(\infty),
%\end{equation}

\begin{widetext}

\begin{figure*}[htbp] 
\centering

% ================= 1. 共享的 Y 轴标题 (放在最左侧) =================
% 使用 \rotatebox 旋转 90 度，\Large 调大字体（根据你的偏好可改）
% 用 \hspace 把它往左边推一点，用 \vspace 微调它的垂直高度
\hspace*{-0.1cm}\raisebox{1.5cm}{\rotatebox{90}{\(\mathrm{V(S_i)}\ (\mathrm{MeV})\)}}
\hspace{-0.1cm}%
% ================= 2. 横向排列的 5 张图 =================
\begin{subfigure}{0.19\textwidth}
\centering
\makebox[\linewidth][c]{%
  \includegraphics[width=\linewidth]{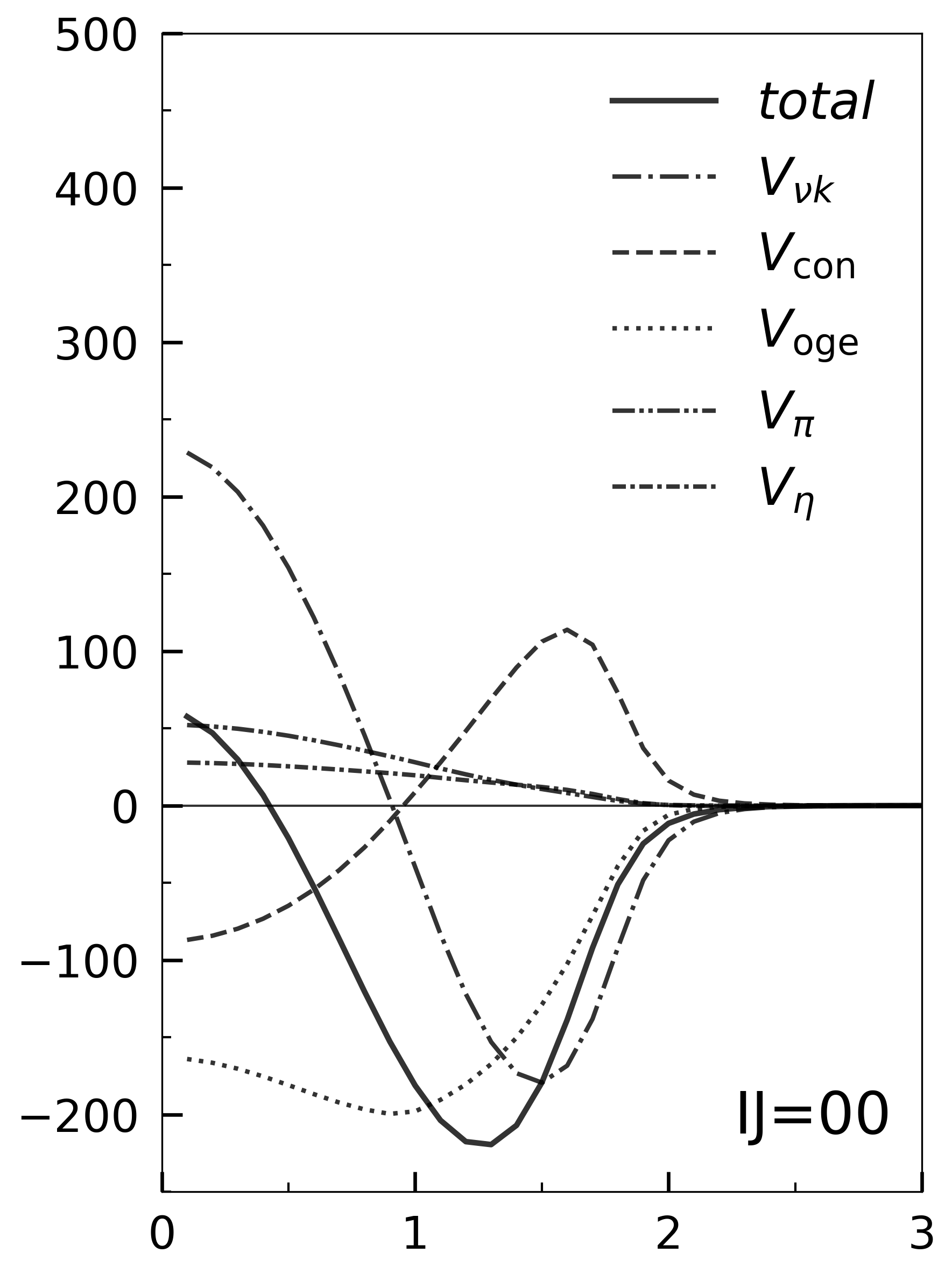}%
  \hspace*{-\linewidth}%
  \makebox[\linewidth][l]{\raisebox{13pt}{\hspace{21pt}\scriptsize (a)}}%
}
\end{subfigure}%
\hspace{-0.3em}
\begin{subfigure}{0.19\textwidth}
\centering
\makebox[\linewidth][c]{%
  \includegraphics[width=\linewidth]{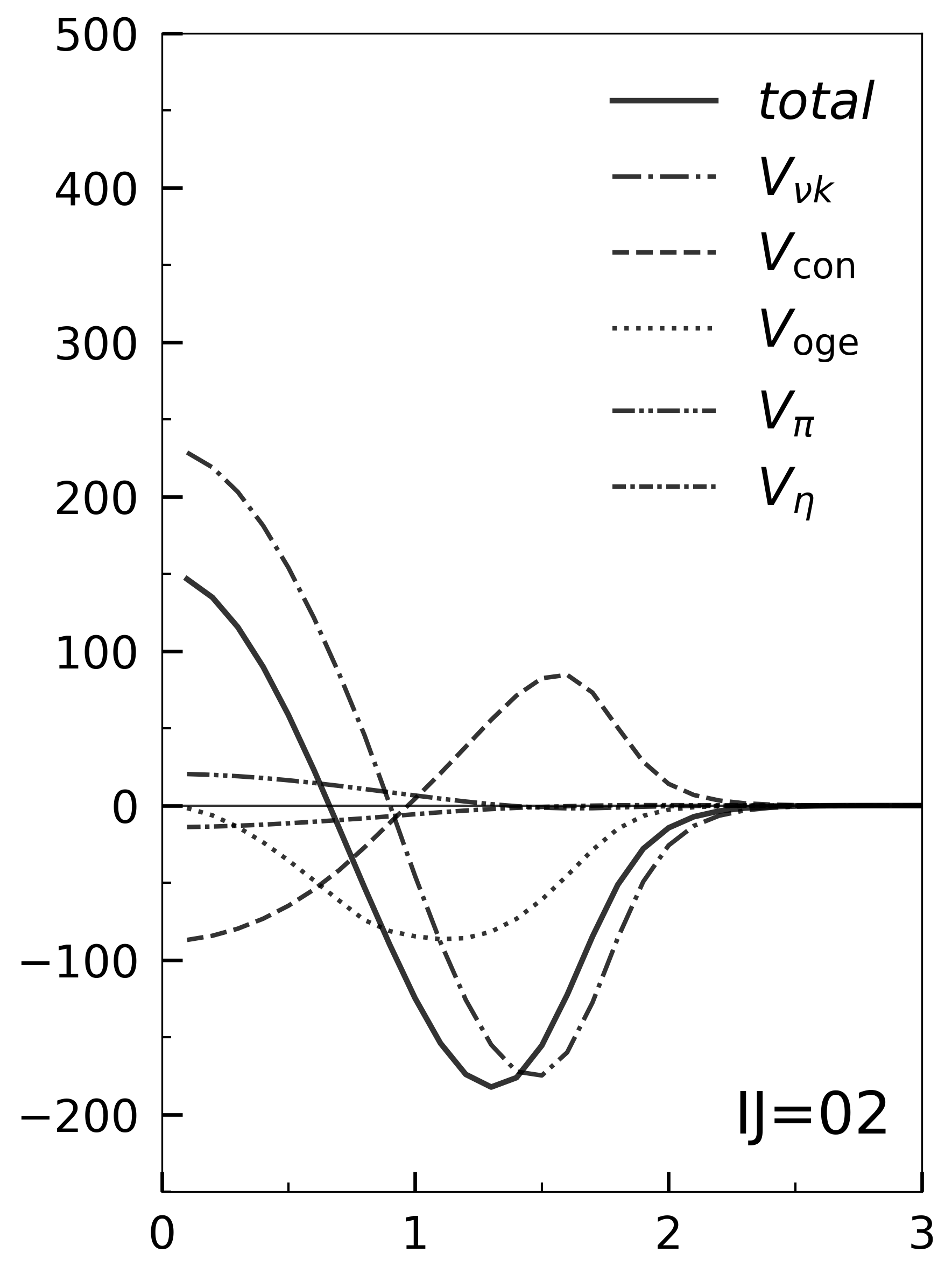}%
  \hspace*{-\linewidth}%
  \makebox[\linewidth][l]{\raisebox{13pt}{\hspace{21pt}\scriptsize (b)}}%
}
\end{subfigure}%
\hspace{-0.3em}
\begin{subfigure}{0.19\textwidth}
\centering
\makebox[\linewidth][c]{%
  \includegraphics[width=\linewidth]{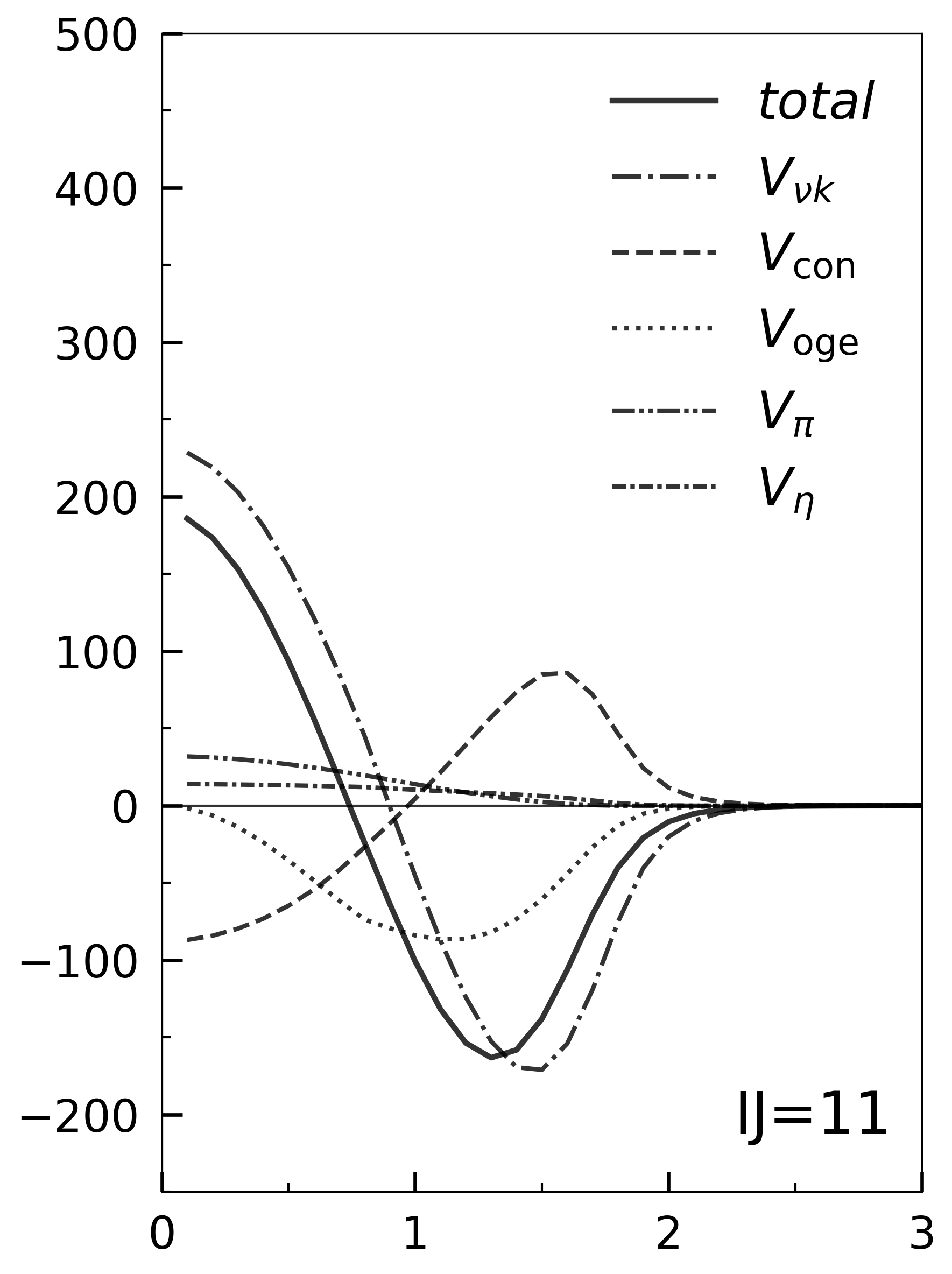}%
  \hspace*{-\linewidth}%
  \makebox[\linewidth][l]{\raisebox{13pt}{\hspace{21pt}\scriptsize (c)}}%
}
\end{subfigure}%
\hspace{-0.3em}
\begin{subfigure}{0.19\textwidth}
\centering
\makebox[\linewidth][c]{%
  \includegraphics[width=\linewidth]{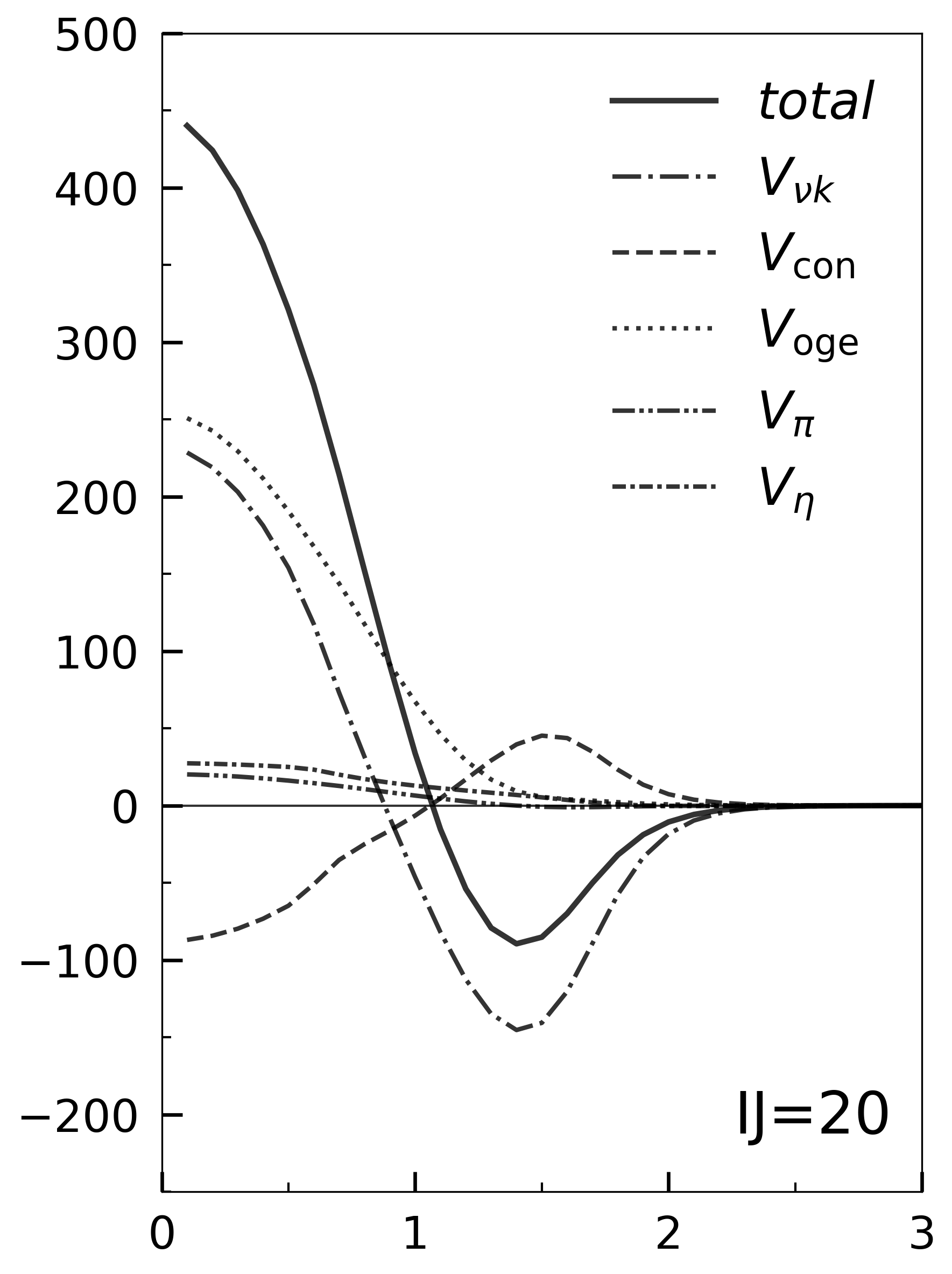}%
  \hspace*{-\linewidth}%
  \makebox[\linewidth][l]{\raisebox{13pt}{\hspace{21pt}\scriptsize (d)}}%
}
\end{subfigure}%
\hspace{-0.3em}
\begin{subfigure}{0.19\textwidth}
\centering
\makebox[\linewidth][c]{%
  \includegraphics[width=\linewidth]{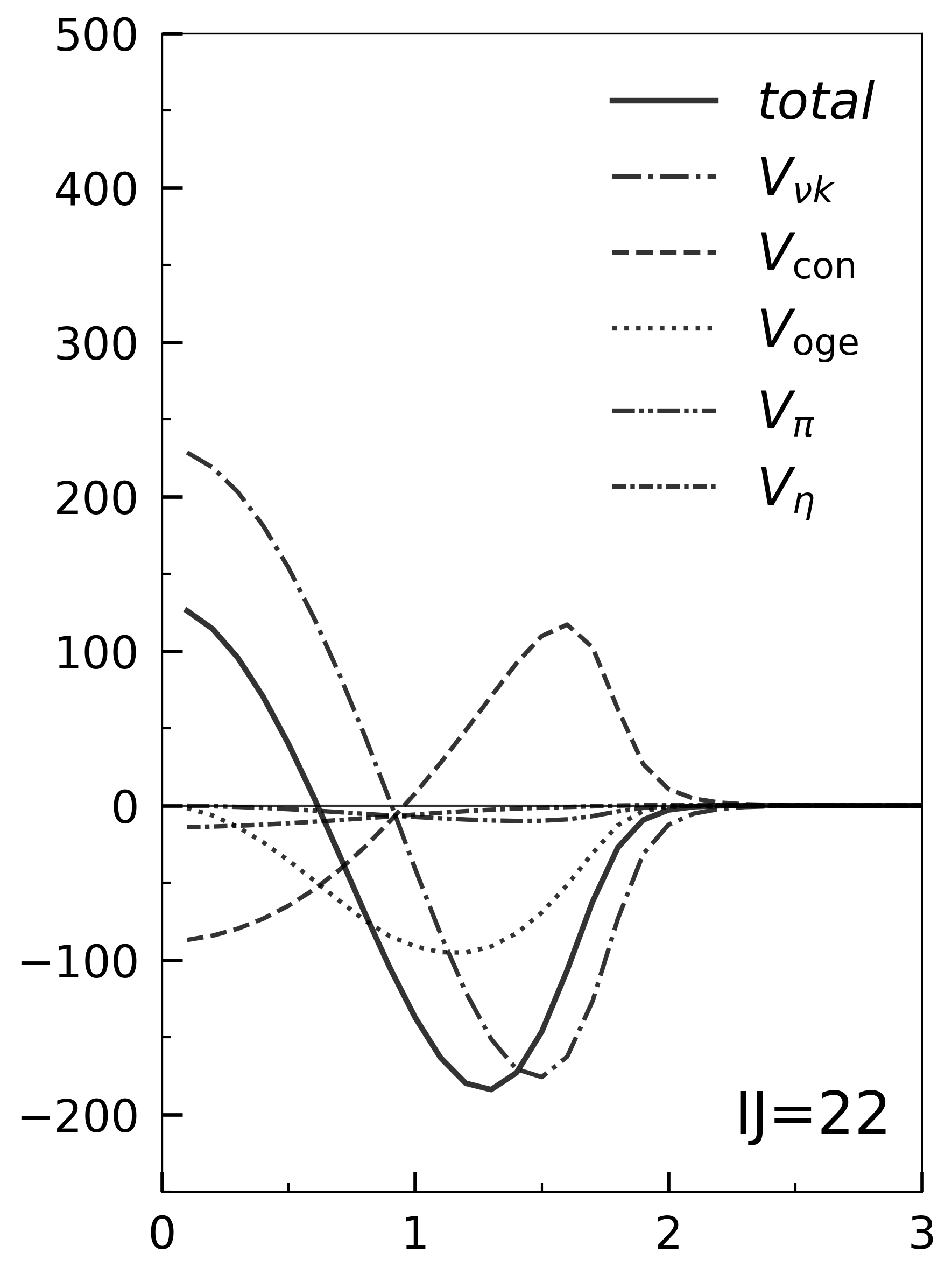}%
  \hspace*{-\linewidth}%
  \makebox[\linewidth][l]{\raisebox{13pt}{\hspace{21pt}\scriptsize (e)}}%
}
\end{subfigure}

% ================= 3. 共享的 X 轴标题 (放在最下方居中) =================
% 使用 \\ 换行，并用 \vspace 调整它与上方图片的距离
\vspace{1ex}
{\(\mathrm{S_i}\ (\mathrm{fm})\)}

\caption{Figure (a) to (e) are the contributions to the effective potential of $\rho\rho$ interactions with different \(IJ\).}
\label{fig:potentials}
\end{figure*}
\end{widetext}

The total effective potential is attractive for all possible $IJ$, with $IJ=00$ showing a visibly large attraction and $IJ=20$ the weakest. This is consistent with the magnitude of the binding energies given in Table~\ref{binding}. The kinetic energy provides the dominant intermediate-range attraction, which is related to the QDCSM employed here. In the QDCSM model, the quark delocalization effect permits quarks to move between two clusters, reducing the total kinetic energy of the system and thereby allowing the kinetic energy to contribute attraction. The delocalization parameter is determined by the dynamics of the system and the results are shown in Fig.~2.

The delocalization parameter for $IJ=00$ is close to $1$ when the distance $S_i$ is less than $0.9$ fm, and that for $IJ=11$ becomes close to $1$ when $S_i$ $\leq$ $0.7$ fm. This indicates that the quarks tend to move between two clusters, giving rise to an attractive interaction. 
For both $IJ=02$ and $IJ=22$, the parameters are close to $1$ when $S_i$ $\leq$ $0.8$ fm, but the former approaches to $0$ more quickly as $S_i$ increases. For $IJ=20$, however, the delocalization parameter approaches to $1$ only when $S_i$ $\leq$ $0.5$ fm and it approaches to 0 quickly as the distance increases. Although this also reduces the kinetic energy, the reduction is much smaller than that in other $IJ$ states, accounting for the weakest attraction of the $IJ=20$ state.

\begin{figure}[htbp]
    \centering
    \hspace*{-0.5cm} 
    \begin{minipage}{1.1\linewidth}
        \centering
        \raisebox{0.35\height}{\rotatebox{90}{\(\varepsilon(S_i)\)}}%
        \hspace{0.2cm}
        \begin{minipage}{0.8\linewidth}
            \centering
            \includegraphics[width=\linewidth]{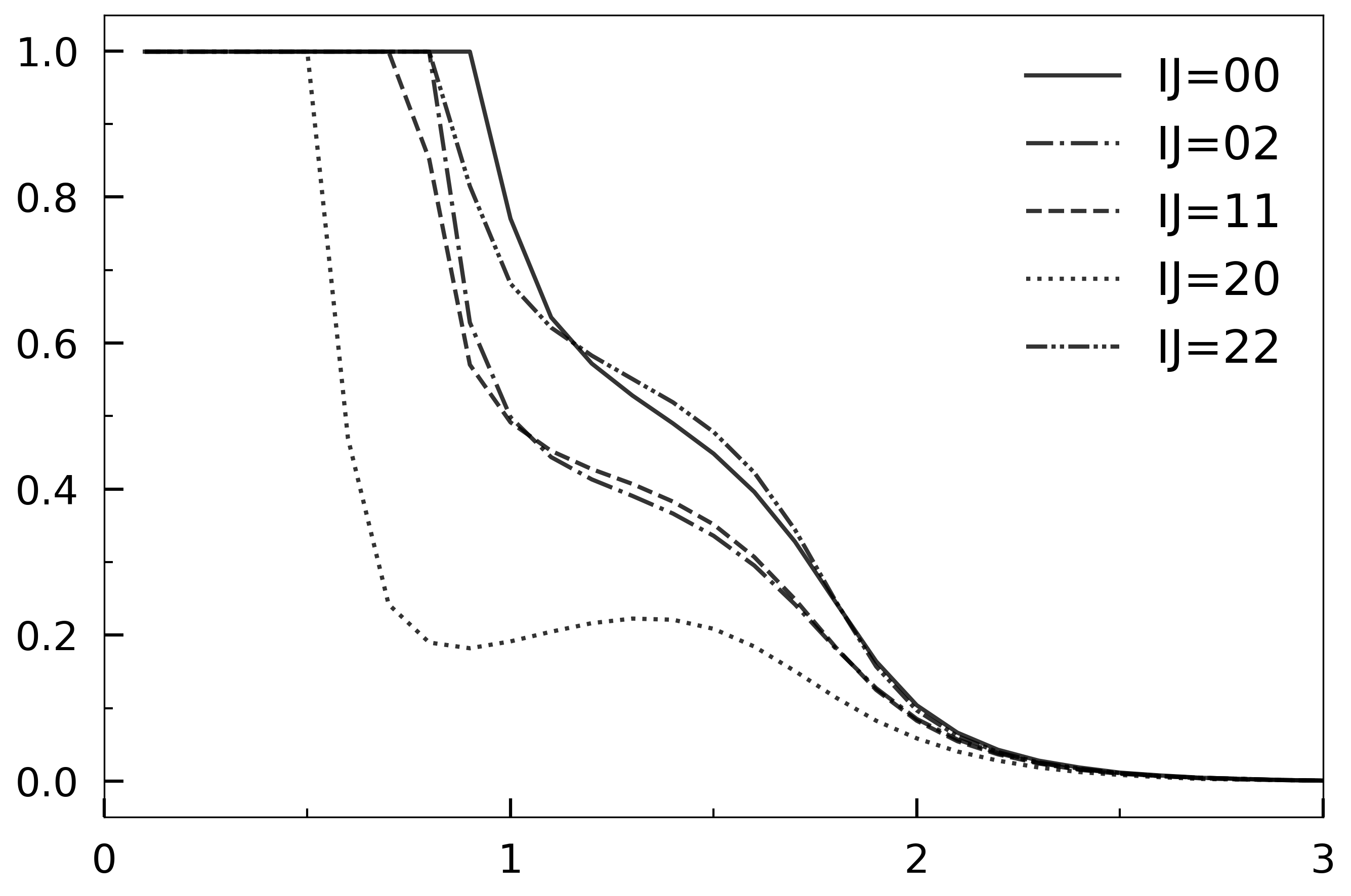}
            
            \vspace{-0.8ex}
            {\(\mathrm{S_i}\ (\mathrm{fm})\)}
        \end{minipage}
    \end{minipage}
    
    \caption{The delocalization parameter $\varepsilon(S_i)$ for $\rho\rho$.}
    \label{fig:delocalizationrho}
\end{figure}

Back to Fig.~1, the confinement potential is attractive at short distances but gives a repulsive contribution in the intermediate distance region for all $IJ$ states. As for the OGE potential, it is purely repulsive only for $IJ=20$ and attractive in other channels. Besides the contribution of other items is small.

To further investigate the origin of the attraction or repulsion from the effective potentials, especially the OGE potential, we calculate the matrix elements in spin-flavor-color space. The definitions are given as follows:
\begin{eqnarray}
N_D & = & \langle \psi | \psi \rangle, \\
{N_{E_1}} & = & -\langle \psi | P_{13} | \psi \rangle, \\
{N_{E_2}} & = & -\langle \psi | P_{24} | \psi \rangle, \\
{N_{E_3}} & = & \langle \psi | P_{13}P_{24} | \psi \rangle, \\
\langle H_{ij}^c \rangle_D & = & \langle \psi | \boldsymbol{\lambda}_i^c \cdot \boldsymbol{\lambda}_j^c | \psi \rangle, \\
\langle H_{ij}^c \rangle_{E_1} & = & -\langle \psi | \boldsymbol{\lambda}_i^c \cdot \boldsymbol{\lambda}_j^c P_{13} | \psi \rangle, \\
\langle H_{ij}^c \rangle_{E_2} & = & -\langle \psi | \boldsymbol{\lambda}_i^c \cdot \boldsymbol{\lambda}_j^c P_{24} | \psi \rangle, \\
\langle H_{ij}^c \rangle_{E_3} & = & \langle \psi | \boldsymbol{\lambda}_i^c \cdot \boldsymbol{\lambda}_j^c P_{13}P_{24} | \psi \rangle, \\
\langle H_{ij}^{cs} \rangle_D & = & \langle \psi | \boldsymbol{\lambda}_i^c \cdot \boldsymbol{\lambda}_j^c \boldsymbol{\sigma}_i \cdot \boldsymbol{\sigma}_j | \psi \rangle, \\
\langle H_{ij}^{cs} \rangle_{E_1} & = & -\langle \psi | \boldsymbol{\lambda}_i^c \cdot \boldsymbol{\lambda}_j^c \boldsymbol{\sigma}_i \cdot \boldsymbol{\sigma}_j P_{13} | \psi \rangle, \\
\langle H_{ij}^{cs} \rangle_{E_2} & = & -\langle \psi | \boldsymbol{\lambda}_i^c \cdot \boldsymbol{\lambda}_j^c \boldsymbol{\sigma}_i \cdot \boldsymbol{\sigma}_j P_{24} | \psi \rangle, \\
\langle H_{ij}^{cs} \rangle_{E_3} & = & \langle \psi | \boldsymbol{\lambda}_i^c \cdot \boldsymbol{\lambda}_j^c \boldsymbol{\sigma}_i \cdot \boldsymbol{\sigma}_j P_{13}P_{24} | \psi \rangle.
\end{eqnarray}

Where $i$ and $j$ represent the $i$-th and $j$-th quarks, respectively. The subscript $D$ means the direct matrix elements and $E$ denotes the exchange matrix elements. According to the anti-symmetrization operator $\mathcal{A}~(\mathcal{A}=1 - P_{13} - P_{24} + P_{13}P_{24})$, there are three exchange terms labeled by the subscripts $1$, $2$ and $3$ in matrix elements $N_{E_1}$, $N_{E_2}$ and $N_{E_3}$. 
The matrix elements $\langle H_{ij}^c \rangle$ represent the color-related factor of the interaction, including the confinement term and the Coulomb part of the one-gluon exchange interaction. The matrix elements $\langle H_{ij}^{cs} \rangle$ mainly describe the spin-color factor associated with the chromomagnetic part of the OGE potential. The calculated matrix elements for all $IJ$ channels are listed in Tables~\ref{matrix00}-\ref{matrix22}.

\begin{table}[htbp]
    \centering
    % 将 5 个表格作为一个小整体，放在一个 minipage 里
    \begin{minipage}{\linewidth}
        \centering
        % ===== 表 1 =====
        \captionof{table}{The matrix elements in spin-flavor-color-isospin spaces of \(\rho\rho\) with \(IJ = 00\).}
        \label{matrix00}
        \renewcommand{\arraystretch}{1.55}
        \setlength{\tabcolsep}{3pt}
        \resizebox{\linewidth}{!}{%
          \begin{tabular}{ccccccccc}
            \hline\hline
            \(ij\) & \(\langle H_{ij}^c \rangle_D\) & \(\langle H_{ij}^c \rangle_{E_1}\) & \(\langle H_{ij}^c \rangle_{E_2}\) & \(\langle H_{ij}^c \rangle_{E_3}\) & \(\langle H_{ij}^{cs} \rangle_D\) & \(\langle H_{ij}^{cs} \rangle_{E_1}\) & \(\langle H_{ij}^{cs} \rangle_{E_2}\) & \(\langle H_{ij}^{cs} \rangle_{E_3}\) \\
            \hline
            12 & \(-16/3\) & 4/9 & 4/9 & \(-16/3\) & \(-16/3\) & 4/9 & 4/9 & \(-16/3\) \\
            13 & 0 & \(-4/9\) & \(-4/9\) & 0 & 0 & 20/9 & 20/9 & 0 \\
            14 & 0 & 4/9 & 4/9 & 0 & 0 & 4/9 & 4/9 & 0 \\
            23 & 0 & 4/9 & 4/9 & 0 & 0 & 4/9 & 4/9 & 0 \\
            24 & 0 & \(-4/9\) & \(-4/9\) & 0 & 0 & 20/9 & 20/9 & 0 \\
            34 & \(-16/3\) & 4/9 & 4/9 & \(-16/3\) & \(-16/3\) & 4/9 & 4/9 & \(-16/3\) \\
            \hline\hline
          \end{tabular}
        }
        \vspace{12pt} % 控制表1和表2的间距
        
        % ===== 表 2 =====
        \captionof{table}{The matrix elements in spin-flavor-color-isospin spaces of \(\rho\rho\) with \(IJ = 02\).}
        \label{matrix02}
        \renewcommand{\arraystretch}{1.55}
        \setlength{\tabcolsep}{3pt}
        \resizebox{\linewidth}{!}{%
          \begin{tabular}{ccccccccc}
            \hline\hline
            \(ij\) & \(\langle H_{ij}^c \rangle_D\) & \(\langle H_{ij}^c \rangle_{E_1}\) & \(\langle H_{ij}^c \rangle_{E_2}\) & \(\langle H_{ij}^c \rangle_{E_3}\) & \(\langle H_{ij}^{cs} \rangle_D\) & \(\langle H_{ij}^{cs} \rangle_{E_1}\) & \(\langle H_{ij}^{cs} \rangle_{E_2}\) & \(\langle H_{ij}^{cs} \rangle_{E_3}\) \\
            \hline
            12 & \(-16/3\) & \(-8/9\) & \(-8/9\) & \(-16/3\) & \(-16/3\) & \(-8/9\) & \(-8/9\) & \(-16/3\) \\
            13 & 0 & 8/9 & 8/9 & 0 & 0 & 8/9 & 8/9 & 0 \\
            14 & 0 & \(-8/9\) & \(-8/9\) & 0 & 0 & \(-8/9\) & \(-8/9\) & 0 \\
            23 & 0 & \(-8/9\) & \(-8/9\) & 0 & 0 & \(-8/9\) & \(-8/9\) & 0 \\
            24 & 0 & 8/9 & 8/9 & 0 & 0 & 8/9 & 8/9 & 0 \\
            34 & \(-16/3\) & \(-8/9\) & \(-8/9\) & \(-16/3\) & \(-16/3\) & \(-8/9\) & \(-8/9\) & \(-16/3\) \\
            \hline\hline
          \end{tabular}
        }
        \vspace{12pt}
        
        % ===== 表 3 =====
        \captionof{table}{The matrix elements in spin-flavor-color-isospin spaces of \(\rho\rho\) with \(IJ = 11\).}
        \label{matrix11}
        \renewcommand{\arraystretch}{1.55} 
        \setlength{\tabcolsep}{3pt} 
        \resizebox{\linewidth}{!}{%
          \begin{tabular}{ccccccccc}
            \hline\hline
            \(ij\) & \(\langle H_{ij}^c \rangle_D\) & \(\langle H_{ij}^c \rangle_{E_1}\) & \(\langle H_{ij}^c \rangle_{E_2}\) & \(\langle H_{ij}^c \rangle_{E_3}\) & \(\langle H_{ij}^{cs} \rangle_D\) & \(\langle H_{ij}^{cs} \rangle_{E_1}\) & \(\langle H_{ij}^{cs} \rangle_{E_2}\) & \(\langle H_{ij}^{cs} \rangle_{E_3}\) \\
            \hline
            12 & \(-16/3\) & 0 & 0 & \(-16/3\) & \(-16/3\) & 0 & 0 & \(-16/3\) \\
            13 & 0 & 0 & 0 & 0 & 0 & 0 & 0 & 0 \\
            14 & 0 & 0 & 0 & 0 & 0 & 0 & 0 & 0 \\
            23 & 0 & 0 & 0 & 0 & 0 & 0 & 0 & 0 \\
            24 & 0 & 0 & 0 & 0 & 0 & 0 & 0 & 0 \\
            34 & \(-16/3\) & 0 & 0 & \(-16/3\) & \(-16/3\) & 0 & 0 & \(-16/3\) \\
            \hline\hline
          \end{tabular}
        }
        \vspace{12pt}
        
        % ===== 表 4 =====
        \captionof{table}{The matrix elements in spin-flavor-color-isospin spaces of \(\rho\rho\) with \(IJ = 20\).}
        \label{matrix20}
        \renewcommand{\arraystretch}{1.55} 
        \setlength{\tabcolsep}{3pt} 
        \resizebox{\linewidth}{!}{%
          \begin{tabular}{ccccccccc}
            \hline\hline
            \(ij\) & \(\langle H_{ij}^c \rangle_D\) & \(\langle H_{ij}^c \rangle_{E_1}\) & \(\langle H_{ij}^c \rangle_{E_2}\) & \(\langle H_{ij}^c \rangle_{E_3}\) & \(\langle H_{ij}^{cs} \rangle_D\) & \(\langle H_{ij}^{cs} \rangle_{E_1}\) & \(\langle H_{ij}^{cs} \rangle_{E_2}\) & \(\langle H_{ij}^{cs} \rangle_{E_3}\) \\
            \hline
            12 & \(-16/3\) & \(-8/9\) & \(-8/9\) & \(-16/3\) & \(-16/3\) & \(-8/9\) & \(-8/9\) & \(-16/3\) \\
            13 & 0 & 8/9 & 8/9 & 0 & 0 & \(-40/9\) & \(-40/9\) & 0 \\
            14 & 0 & \(-8/9\) & \(-8/9\) & 0 & 0 & \(-8/9\) & \(-8/9\) & 0 \\
            23 & 0 & \(-8/9\) & \(-8/9\) & 0 & 0 & \(-8/9\) & \(-8/9\) & 0 \\
            24 & 0 & 8/9 & 8/9 & 0 & 0 & \(-40/9\) & \(-40/9\) & 0 \\
            34 & \(-16/3\) & \(-8/9\) & \(-8/9\) & \(-16/3\) & \(-16/3\) & \(-8/9\) & \(-8/9\) & \(-16/3\) \\
            \hline\hline
          \end{tabular}
        }
        \vspace{12pt}
        
        % ===== 表 5 =====
        \captionof{table}{The matrix elements in spin-flavor-color-isospin spaces of \(\rho\rho\) with \(IJ = 22\).}
        \label{matrix22}
        \renewcommand{\arraystretch}{1.55} 
        \setlength{\tabcolsep}{3pt} 
        \resizebox{\linewidth}{!}{%
          \begin{tabular}{ccccccccc}
            \hline\hline
            \(ij\) & \(\langle H_{ij}^c \rangle_D\) & \(\langle H_{ij}^c \rangle_{E_1}\) & \(\langle H_{ij}^c \rangle_{E_2}\) & \(\langle H_{ij}^c \rangle_{E_3}\) & \(\langle H_{ij}^{cs} \rangle_D\) & \(\langle H_{ij}^{cs} \rangle_{E_1}\) & \(\langle H_{ij}^{cs} \rangle_{E_2}\) & \(\langle H_{ij}^{cs} \rangle_{E_3}\) \\
            \hline
            12 & 16/9 & 16/9 & \(-8/9\) & \(-16/3\) & \(-16/3\) & 16/9 & 16/9 & \(-16/3\) \\
            13 & 0 & \(-16/9\) & \(-16/9\) & 0 & 0 & \(-16/9\) & \(-16/9\) & 0 \\
            14 & 0 & 16/9 & 16/9 & 0 & 0 & 16/9 & 16/9 & 0 \\
            23 & 0 & 16/9 & 16/9 & 0 & 0 & 16/9 & 16/9 & 0 \\
            24 & 0 & \(-16/9\) & \(-16/9\) & 0 & 0 & \(-16/9\) & \(-16/9\) & 0 \\
            34 & \(-16/3\) & 16/9 & 16/9 & \(-16/3\) & \(-16/3\) & 16/9 & 16/9 & \(-16/3\) \\
            \hline\hline
          \end{tabular}
        }
    \end{minipage}
\end{table}

The matrix elements $\langle H_{ij}^c \rangle$ between the two clusters are zero as presented in Tables~\ref{matrix00}-\ref{matrix22}. This is because both meson clusters are color-singlets, and do not contribute to the meson-meson interaction. The behavior of the matrix elements $\langle H_{ij}^{cs} \rangle$ is consistent with the variation of the effective potential, as illustrated by the two notable cases $IJ=00$ and $IJ=20$. For $IJ=00$, the exchange matrix elements $\langle H_{ij}^{cs} \rangle_{E_1}$ and $\langle H_{ij}^{cs} \rangle_{E_2}$ are all positive; when the negative sign of the color-magnetic term in the $V^{OGE}_{ij}$ is considered, the resulting values are negative, which confirms the deep attraction of the OGE potential. In case of $IJ=20$, the exchange matrix elements $\langle H_{ij}^{cs} \rangle_{E_1}$ and $\langle H_{ij}^{cs} \rangle_{E_2}$ are all negative. With the sign taken into account, the results become positive, corresponding to the repulsive effect of the OGE potential shown in Fig.~1.

In addition, to further explore the similarities between the vector-vector meson interaction and the decuplet-decuplet baryon interaction, the interaction between two decuplet baryons is examined. Taking $\Delta\Delta$ as an example, a previous study conducted in the QDCSM shows that the $IJ=03$ state has the binding energy of $-95$ MeV, while the state with $IJ=30$ is weakly bound with the value of only $-30$ MeV~\cite{Huang}. To reveal the underlying dynamics, the contributions to the effective potential of $\Delta\Delta$ interaction with $IJ=03$ and $IJ=30$ are shown in Fig.~3.

\begin{figure}[htbp] 
\centering
\hspace*{-0.1cm}\raisebox{1.8cm}{\rotatebox{90}{\(\mathrm{V(S_i)}\ (\mathrm{MeV})\)}}
\hspace{-0.1cm}%
\begin{subfigure}{0.22\textwidth}
\centering
\makebox[\linewidth][c]{%
  \includegraphics[width=\linewidth]{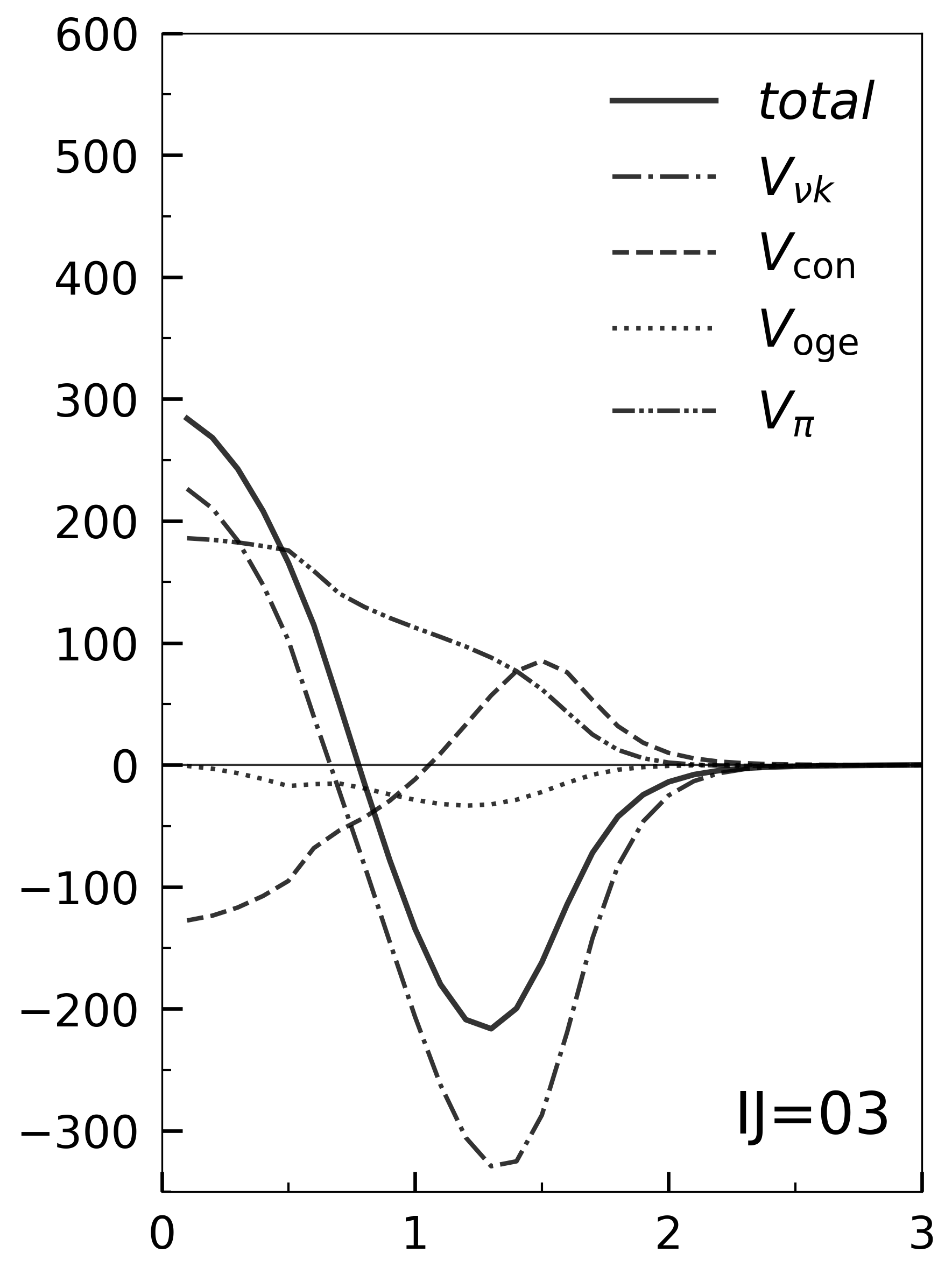}%
  \hspace*{-\linewidth}%
  \makebox[\linewidth][l]{\raisebox{15pt}{\hspace{25pt}\scriptsize (a)}}%
}
\end{subfigure}%
\hspace{-0.1em}
\begin{subfigure}{0.22\textwidth}
\centering
\makebox[\linewidth][c]{%
  \includegraphics[width=\linewidth]{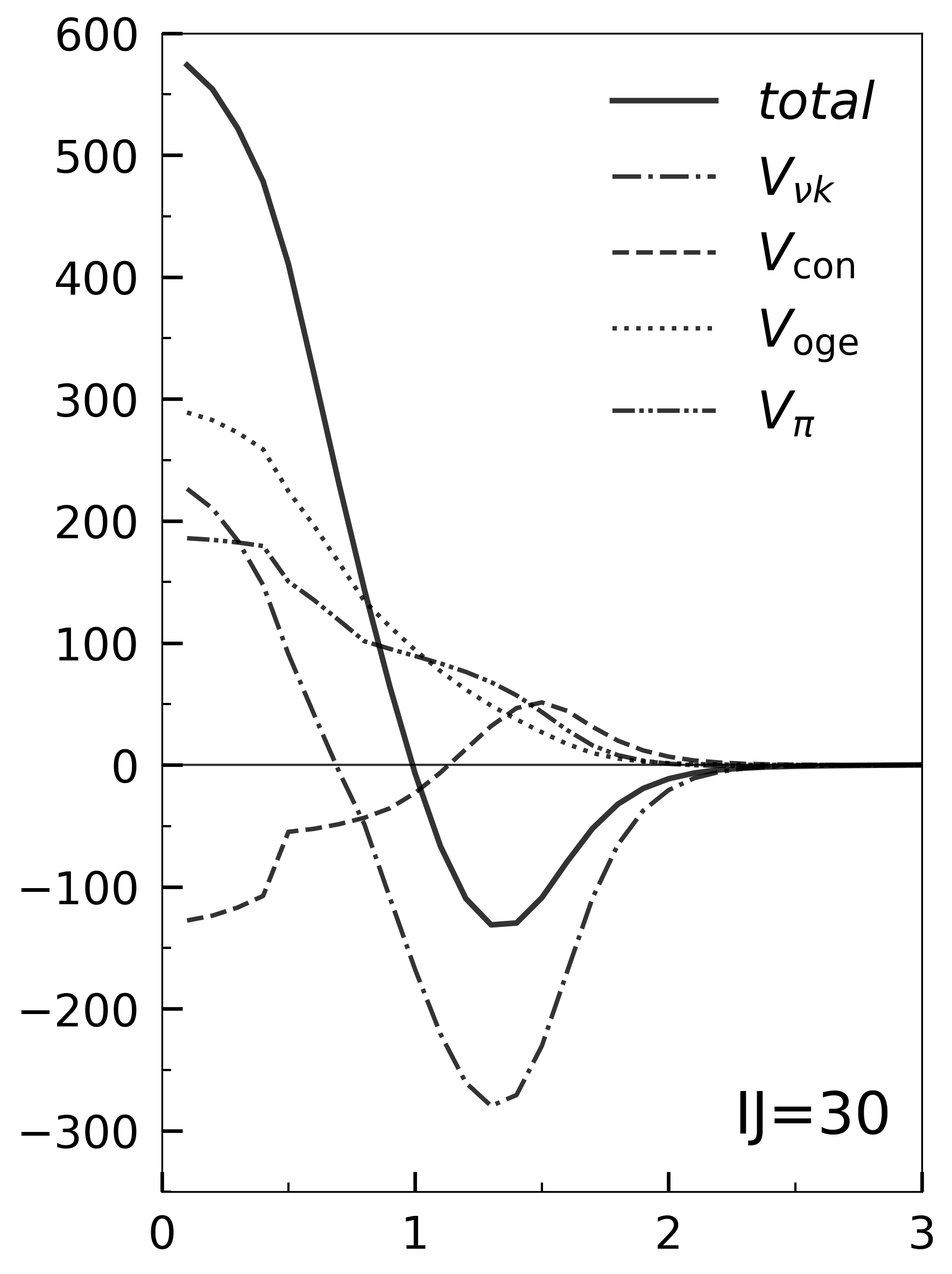}%
  \hspace*{-\linewidth}%
  \makebox[\linewidth][l]{\raisebox{15pt}{\hspace{25pt}\scriptsize (b)}}%
}
\end{subfigure}
\vspace{1ex}
{\(\mathrm{S_i}\ (\mathrm{fm})\)}

\caption{The contributions to the effective potential of $\Delta\Delta$ interactions with $IJ=03$ and $IJ=30$.}
\label{fig:d*}
\end{figure}

\begin{figure}[htbp]
    \centering
    \hspace*{-0.5cm}
    \begin{minipage}{1.1\linewidth}
        \centering
        \raisebox{0.3\height}{\rotatebox{90}{$\varepsilon(S_i)$}}%
        \hspace{0.2cm}
        \begin{minipage}{0.8\linewidth}
            \centering
            \includegraphics[width=\linewidth]{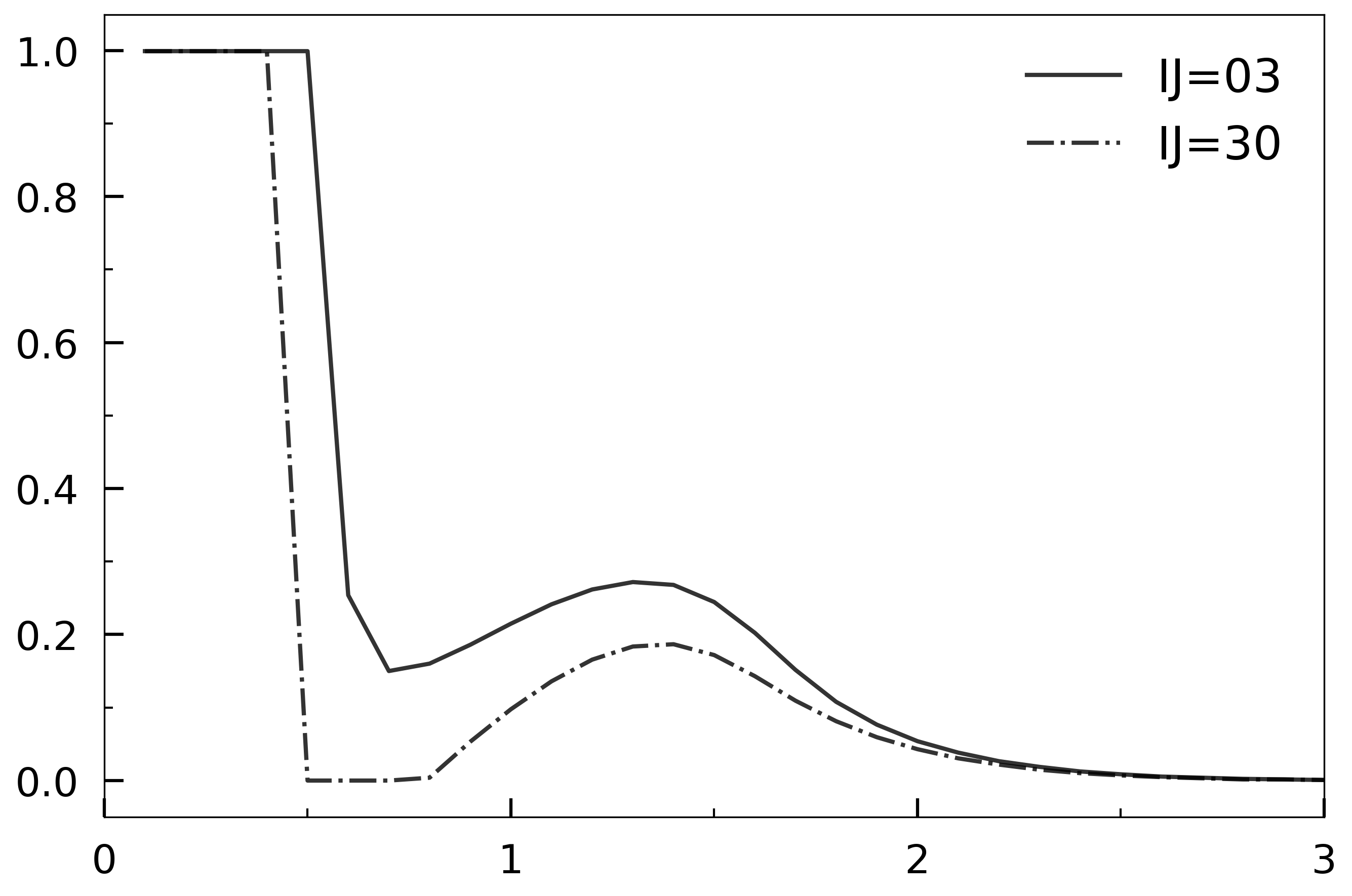}
            
            \vspace{-0.8ex}
            {\(\mathrm{S_i}\ (\mathrm{fm})\)}
        \end{minipage}
    \end{minipage}
    
    \caption{The delocalization parameter $\varepsilon(S_i)$ for $\Delta\Delta$.}
    \label{fig:delocalizationd*}
\end{figure}

For both states, the attraction originates mainly from the kinetic energy term, while the $\pi$ meson exchange potential are repulsive, similar to the behavior found for $\rho\rho$ states. The delocalization parameter for $IJ=03$ approaches to 1 when the distance $S_i$ is less than $0.5$ fm, while that for $IJ=30$ is close to 1 when $S_i$ $\leq$ $0.4$ fm and it quickly approaches to $0$. This accounts for the relatively weaker kinetic attraction in $IJ=30$ compared with $IJ=03$. 
At short range, the confinement potential is attractive, whereas at intermediate range it becomes repulsive. The OGE potential provides attraction for $IJ=03$, while it is repulsive for $IJ=30$. As a result, the total attraction for $IJ=03$ is larger than that for $IJ=30$. 
The properties of the interaction between two $\rho$ mesons are similar to those of the interaction between two $\Delta$ baryons. Since the dibaryon with $IJ=03$ composed of two $\Delta$s has been predicted as an ``inevitable" nonstrange state~\cite{d*} and confirmed by experiments~\cite{WASA08,WASA11,WASA14}, it is natural to consider whether the $\rho\rho$ states can likewise be identified as ``inevitable'' nonstrange tetraquark states. 

\begin{table}[htbp]
\centering
\caption{Decay channels for \(\rho\rho\) with different \(IJ\).}
\label{decay_channels}
\renewcommand{\arraystretch}{1.6} % 整体行高稍微放大，让表格透气
\setlength{\tabcolsep}{8pt}       % 默认列间距
% 核心修改：全部使用基础的 c 列居中
\begin{tabular}{c c c}
\hline \hline
\(IJ\) & \(S\)-wave & \(D\)-wave \\ \hline

% IJ=00 紧凑排列，由于没有宽度限制，内容会非常紧凑
\(00\) & \(\pi\pi\), \(\eta\eta\), \(KK\) & \\

\(02\) &  & \(\pi\pi\), \(\pi\rho\), \(\eta\eta\), \(\eta\omega\), \(KK\), \(KK^*\)\\

\(11\) & \(\pi\rho\), \(\pi\omega\), \(\pi\phi\), \(\eta\rho\), \(KK^*\) & \(\pi\rho\), \(\pi\omega\), \(\pi\phi\), \(\eta\rho\), \(KK^*\) \\
   
\(20\) & \(\pi\pi\) & \\

\(22\) &  & \(\pi\pi\) \\
\hline \hline
\end{tabular}
\end{table}

%Given the deepest binding energy of the $\rho\rho$ states with $IJ=00$ obtained in this work, this state seems to be the promising candidate of the ``inevitable'' tetraquark states. 
A preliminary analysis of the $\rho\rho$ decay channels for different $IJ$ is carried out, so as to provide better predictions. Table~\ref{decay_channels} shows that the $IJ=11$ state can decay into $S$- and $D$-wave channels while the states with $IJ=00$ and $IJ=20$ can only decay into $S$-wave channels. For example, both $IJ=00$ and $IJ=20$ $\rho\rho$ can decay into $S$-wave $\pi\pi$. Given that the $\rho\rho$ states considered here are all in the $S$-wave, the relevant decay is an $S$-to-$S$-wave one, mediated by the central-force coupling. Because this coupling is relatively strong, two possible cases may arise. In one case, the coupling may push the energy of the state above the $\rho\rho$ threshold, turning it into a scattering state. In the other case, $\rho\rho$ may become a resonance after coupling, but with a large decay width that makes the resonance very broad and highly unstable. 
This is analogous to the coupling of the $\Delta\Delta$ $(IJ=01)$ to the $NN$. As studied in Ref.~\cite{Ping09}, the $\Delta\Delta$ with $IJ=01$ has a relatively deep binding energy of $258$ MeV. However, after coupling to the $S$-wave $NN$ channel, its energy is pushed above the threshold, causing the resonance to disappear. 
In contrast, the $\rho\rho$ states with $IJ=02$ and $IJ=22$ do not decay into $S$-wave channels, but only into $D$-wave channels. Generally, the $S$-to-$D$-wave decay is driven by the tensor-force coupling, which is relatively weak compared with the central-force coupling. Consequently, both $IJ=02$ and $IJ=22$ states are less affected by channel coupling. This is similar to the $\Delta\Delta$ state with $IJ=03$, which can decay to $D$-wave $NN$. After the consideration of channel coupling, the $\Delta\Delta$ state appears as a resonance state in the $D$-wave $NN$ scattering process and the energy shifts from $2365$ MeV to $2368$ MeV~\cite{Ping09}. 
Based on the above analyse, the $\rho\rho$ states with $IJ=00$, $IJ=11$ and $IJ=20$ might turn into a scattering state through the $S$-wave channels, whereas the states with $IJ=02$ and $IJ=22$ are likely to appear as a resonance in the $D$-wave scattering. Therefore, we tend to consider that $\rho\rho$ with $IJ=02$ and $IJ=22$ may be the ``inevitable'' nonstrange tetraquark states.

\section{Summary}
In this work, we systematically investigate the $S$-wave $\rho\rho$ states with $IJ = 00,~02,~11,~20$ and $22$ in the framework of the QDCSM by the RGM. To confirm the existence of bound states, we calculate the binding energies for all states. The distance between the two clusters is also calculated to explore the structure of $\rho\rho$ states. In order to study the origin of $\rho\rho$ interaction, we investigate the contributions to the effective potential from varies terms of $\rho\rho$ interaction and calculate the matrix elements in spin-flavor-color space. 

The numerical results show that all channels can form bound states, with binding energies of $-124.93$ MeV, $-87.43$ MeV, $-69.64$ MeV, $-2.31$ MeV, and $-90.42$ MeV for $IJ = 00,~02,~11,~20$ and $22$, respectively. Notably, the binding energy for $IJ=00$ is the largest among all states, whereas that for $IJ=20$ is the smallest.  
Based on the calculated distance between the two clusters, we assume that $\rho\rho$ tend to form molecular states.

The study of the contribution of each interaction term shows that the total attraction of $\rho\rho$ comes mainly from the kinetic energy, which is  related to the intermediate-range attraction mechanism in the QDCSM. As the delocalization parameters approaches to 1, the quarks tend to move more strongly between two clusters. This delocalization reduces the total kinetic energy of the system. 
The confinement potential provides an attractive contribution at short distances but a repulsive contribution in the intermediate-range for all states. The OGE potential is attractive, except for $IJ = 20$, where its contribution is repulsive. The contribution of other terms is small.  
The matrix element calculations show that the behavior of the $\langle H_{ij}^{cs} \rangle$ matrix elements is consistent with the variation of the effective potential.%: the exchange matrix elements for $IJ=00$ are negative, confirming the deep OGE attraction, while those for $IJ=20$ are positive, corresponding to its repulsive effect.
 
Additionally, to explore the analogy between the vector-vector meson interaction and the decuplet-decuplet baryon interaction, we take $\Delta\Delta$ as an example to explore the interaction between two decuplet baryons. 
The total attraction for $IJ=03$ is larger than that for $IJ=30$. The behavior of the $\rho\rho$ interaction with different quantum numbers resembles that of the $\Delta\Delta$. 
%As the $\Delta\Delta$ state with $IJ=03$ has been extensively studied and is regarded as an ``inevitable'' state $d^*$, we similarly infer that $\rho\rho$ with $IJ=00$ and $IJ=02$ may be the most promising candidates of the ``inevitable'' tetraquark states.
%However, the $\rho\rho$ tetraquark can decay into $\pi\pi$. 
According to the discussion of the channel coupling effects, we suggest that $\rho\rho$ with $IJ=02$ and $IJ=22$ are ``inevitable" nonstrange tetraquark states, which is worthy of the attention from both theoretical and experimental studies. In future work, we will systematically investigate the scattering processes to confirm whether the $\rho\rho$ states are resonances or not.

\acknowledgments{This work is supported partly by the National Natural Science Foundation of China under Contracts Nos. 12575088, 11675080, 11775118 and 11535005.}

\end{document}